\documentclass[doublecol]{epl2}

\usepackage{graphicx}
\usepackage{dcolumn}
\usepackage{bm}
\usepackage{epsf}
\usepackage{amsmath}
\usepackage{epsfig}

\DeclareSymbolFont{AMSb}{U}{msb}{m}{n}
\DeclareMathSymbol{\N}{\mathbin}{AMSb}{"4E}
\DeclareMathSymbol{\Z}{\mathbin}{AMSb}{"5A}
\DeclareMathSymbol{\R}{\mathbin}{AMSb}{"52}
\DeclareMathSymbol{\Q}{\mathbin}{AMSb}{"51}
\DeclareMathSymbol{\I}{\mathbin}{AMSb}{"49}
\DeclareMathSymbol{\C}{\mathbin}{AMSb}{"43}

\newcommand{\ra}{\rangle}
\newcommand{\be}{\begin{equation}}
\newcommand{\ee}{\end{equation}}
\newcommand{\bea}{\begin{eqnarray}}
\newcommand{\eea}{\end{eqnarray}}

\newcommand{\tr}{\text{Tr}}
\DeclareMathOperator{\arcsinh}{arcsinh}

\title{Energetics of quantum correlations}
\author{Raoul Dillenschneider and Eric Lutz}
\institute{Department of Physics, University of Augsburg, D-86135 Augsburg, Germany}

\abstract{
We consider a photo--Carnot engine that consists of a single--mode radiation field in an optical cavity.
One the heat reservoirs is made of a beam of thermally entangled pairs of two--level atoms that interact resonantly with the cavity. We express the thermodynamic  efficiency of the engine in terms of the quantum discord of the atomic pair and find that it can exceed its classical value. Our results show that useful work can be extracted from  quantum correlations, indicating that the latter are a valuable resource in quantum thermodynamics.}


\begin{document}

\maketitle

\section{Introduction}

Nonclassical correlations are one of the most intriguing consequences 
of quantum theory. A major discovery of the last decade is that the 
efficiency of some computing and information processing  tasks can be 
greatly enhanced when using quantum correlated states \cite{nie00}. 
Quantum correlations are traditionally associated with entanglement, 
that is, nonseparability. However, it has recently been realized that 
the notion of nonclassical correlations is more general 
\cite{ben99,OllivierZurek,hor03}. An information--theoretic measure 
of the quantumness of correlations, the quantum  discord, has been introduced by Ollivier 
and Zurek \cite{OllivierZurek}. It is defined as the difference of two 
expressions of the mutual information that are classically identical 
(see Eq.~\ref{10} below). The quantum discord can be non--zero for states that 
are separable, indicating the presence of quantum correlations in 
nonentangled states.
Remarkably, a recent experiment has demonstrated that states with 
non--zero discord can lead to an exponential speedup in a model of 
quantum computation (deterministic quantum computation with one quantum 
bit -- DQC1), in the {\it absence} of entanglement \cite{dat08,lan08}.


A thermodynamic approach to quantify quantum correlations has been developed in Refs. \cite{opp02,Zurek,hor05,ali04}, exploiting the intimate connection existing between thermodynamics and information theory. The basic quantity in this context is the work deficit, defined as the difference of thermodynamic work that can be extracted globally and locally from a heat bath using a correlated bipartite state. The work deficit has been shown to be   equal to the quantum discord when  one-way communication is allowed \cite{Zurek,hor05}. In this paper, we consider the problem of gaining useful work from quantum correlated states. Specifically, we  discuss  a scheme that permits extraction of thermodynamical work from an ensemble of thermal quantum correlated qubits with non--zero  discord.

We consider the photo--Carnot engine introduced by Scully and 
coworkers \cite{Scully}. The latter consists of a single mode of a quantized 
radiation field inside a resonant high--quality optical cavity. One of the mirrors 
of the cavity (the piston) is driven by the radiation pressure of the thermal 
photons of the field, while the other mirror is used to exchange heat with an 
external reservoir. The coupling to a second (quantum) reservoir is achieved 
by sending individual thermal atoms through the cavity and letting them interact 
with the radiation field. Two cases can then  be distinguished \cite{Scully}: When 
regular two--level atoms are sent through the cavity, the maximum efficiency of 
the quantum engine after a Carnot cycle is found to be given by the classical  efficiency, $\eta_C= 1-T_c/T_h$, where $T_c$ and $T_h$ are the respective 
temperatures of the cold and hot reservoir (which here corresponds to the atomic 
reservoir). The situation changes dramatically when three--level atoms are considered 
instead. By preparing the two (quasi--degenerate) ground states in a coherent 
superposition and by properly tuning the relative phase between the two states, the 
temperature of the cavity can be effectively increased. As a consequence, 
the quantum efficiency can become larger than the classical one, $\eta > \eta_C$. 
Thermodynamic efficiency can therefore be improved by using quantum coherence as a resource.

In order to investigate the effect of quantum correlations in the quantum Carnot 
engine described above, we consider pairwise correlated two--level atoms in a 
thermal entangled state \cite{nie98}. The notable feature of these states is the 
existence of entanglement in thermal equilibrium at finite temperature. Thermal 
entanglement has been extensively studied in solid--state systems, such as various 
Ising or Heisenberg spin models \cite{arn01,lag02,am07}. Signatures of thermal 
entanglement in macroscopic systems have in addition been observed in recent 
experiments \cite{gos03,ver06,rap07}.

In the following, we calculate the thermodynamic efficiency of the photo--Carnot 
engine when a beam of two--level atoms in thermal entangled states is sent through 
the optical cavity. In the limit of very large number of atoms, the beam can be 
considered as a heat reservoir and the efficiency of the quantum engine reduces to the known classical value  when  atoms are not correlated. 
 We treat two different cases: In the first case, the two atoms 
of a correlated pair fly sequentially through the cavity, whereas in the second case, 
only one of the atoms interacts with the radiation field. In both situations, we find 
that the presence of correlations modifies detailed--balance between absorption and 
emission of photons in the cavity and changes it effective temperature; the temperature 
is lowered in the two--atom case, while it is raised in the one--atom case. As a result 
the quantum Carnot engine can outperform the classical engine, showing that quantum correlations are a valuable resource. 

\section{Thermal entanglement}

Let us consider two identical two--level atoms with frequency $\omega$ 
coupled via a XY--Heisenberg interaction of the form,

\begin{eqnarray}
H = \hbar \omega S^z_{1} + \hbar \omega S^z_{2} 
+ \hbar \lambda \left(S^{+}_{1} S^{-}_{2} + S^{-}_{1} S^{+}_{2} 
\right)\ ,
\label{Eq1}
\end{eqnarray}

\noindent
where $\lambda$ is the controllable strength of the interaction. We have 
here defined the spin operators $S^z_j =  \sigma^z_j/2 = \left(|g\rangle \langle g|_j - 
|e\rangle \langle e|_j \right)/2$ and $S^{\pm}_j =  \left(S^x_j \pm i S^y_j \right)$, 
where $\sigma^{x,y,z}$ are the usual Pauli operators. Ground and excited states are 
denoted by $|g\rangle$ and $|e\rangle$. In the context of cavity QED, a scheme to 
entangle in a controlled manner two identical atoms via the XY--Heisenberg interaction 
\eqref{Eq1} has been recently proposed in Ref.~\cite{ZhengGuo} and implemented 
experimentally \cite{osn01}. The state of the two atoms in thermal equilibrium at 
temperature $T$ is described by the density operator, $\rho = Z^{-1} \exp(-\beta H)$, 
where $Z= \tr \exp(-\beta H) = 2 \left(\cosh(\beta \hbar \omega)
+ \cosh(\beta \hbar \lambda) \right)$ is the partition function and $\beta= (kT)^{-1}$.
In the natural basis $\left\{|gg \rangle,|ge \rangle,|eg \rangle,
|ee \rangle \right\}$, the thermal density matrix can be written as,

\begin{eqnarray}
\rho &=& \frac{1}{Z}
\Bigg(
e^{\beta \hbar \omega} |gg\rangle \langle gg| 
+ e^{-\beta \hbar \omega} |ee\rangle \langle ee|
\notag
\\
&&
+ e^{\beta \hbar \lambda} |\Psi_{-} \rangle \langle \Psi_{-}| 
+ e^{-\beta \hbar \lambda} |\Psi_{+}\rangle \langle \Psi_{+}|
\Bigg)\ ,
\label{Eq2}
\end{eqnarray}
where we have introduced the maximally entangled Bell states, 
$|\Psi_{\pm}\rangle =  \left(\pm |ge \rangle + |eg \rangle 
\right)/\sqrt{2}$, which, together with the states $|gg\rangle$ and 
$|ee\rangle$, form an orthogonal basis. It is worth  noticing that the interaction 
parameter $\lambda$ controls the weight of the two Bell states inside the mixture. 
For further reference, we also define the following matrix elements of $\rho$: the matrix 
elements in the ground and excited states, $\rho_g= \langle gg|\rho|gg\rangle= Z^{-1} 
\exp(\beta\hbar \omega)$ and $\rho_e=\langle ee|\rho|ee\rangle 
= Z^{-1} \exp(-\beta\hbar \omega)$, 
as well as the diagonal, $\rho_d=\langle ge|\rho|ge\rangle  = \langle eg|\rho|eg\rangle 
= Z^{-1} \cosh(\beta\hbar \lambda)$ and nondiagonal elements, $\rho_{nd}=\langle 
ge|\rho|eg\rangle = \langle eg|\rho|ge\rangle= -Z^{-1}\sinh(\beta\hbar \lambda) $. 
The nondiagonal matrix elements vanish when $\lambda=0$.

The entanglement properties of the thermal state \eqref{Eq2} have been investigated 
in Ref.~\cite{Wang}. The concurrence can be evaluated in closed form and reads
\be
\mathcal{C} = \max \left\{ 0, 
\frac{\sinh \left( \beta \hbar \lambda \right) - 1}
{\cosh\left(\beta \hbar \omega \right) 
+ \cosh\left(\beta \hbar \lambda \right)} \right\}\ .
\ee
For $\beta \hbar \lambda > \arcsinh(1) \simeq 0.88$, the concurrence 
is non-zero and the two atoms described by Eq.~\eqref{Eq2} are
therefore thermally entangled. However, for $\beta \hbar \lambda \le \arcsinh(1)$, 
the concurrence vanishes and the mixture is separable. The thermal state \eqref{Eq2} 
is thus entangled for sufficiently low temperature and/or strong coupling $\lambda$.

\section{Two atoms through cavity}

\begin{figure}
\center
\epsfig{file=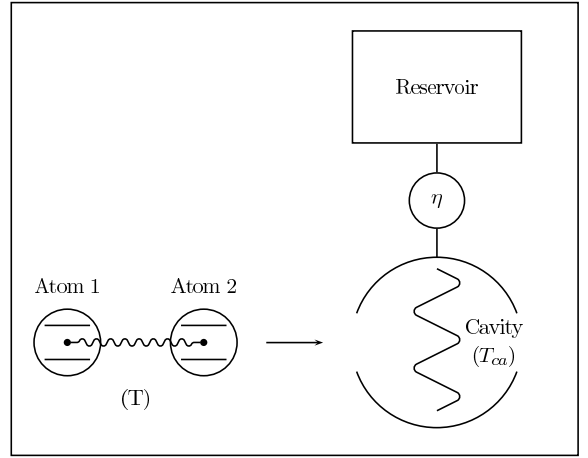,width=8cm}
\caption{Schematic representation of the photo--Carnot engine: A beam of   quantum correlated two--level atoms (quantum reservoir at temperature $T$) is sent  through a resonant optical cavity (system at temperature  $T_{ca}$). Work is produced by putting the cavity cyclically in contact with another reservoir at a different temperature, with a quantum efficiency $\eta$.
}
\label{FigScheme2}
\end{figure}

We now focus on the situation in which atoms of thermal entangled pairs pass through 
the cavity at random time intervals, as depicted in Fig.\ref{FigScheme2}. Each atom, $j=1, 2$, interacts with the single mode 
of the radiation field via a resonant Jaynes--Cummings coupling, $H^{int}_{j} = -g \hbar 
\left( a S_{j}^{+} + a^\dagger S_{j}^{-} \right)$, with coupling constant $g$. The 
Hamiltonian of the cavity mode is $H_{ca} = \hbar \omega a^\dagger a$, where $a$ and 
$a^\dagger$ are the ladder operators of the oscillator.
We are interested in the stationary properties of the single cavity mode described by 
the operator $\rho_{ca}(t)$, when a large number of similarly prepared pairs of atoms
is sent through the cavity. The master equation obeyed by $\rho_{ca}(t)$ can be derived 
in the limit of weak coupling by using standard techniques \cite{Schleich} and reads, 

\begin{eqnarray}
&&\frac{\partial \rho_{ca}}{\partial t} = 
i \omega \left[\rho_{ca},a^\dagger a \right]
\notag \\
&&
- \frac{r_1}{2} \phi^2 \left( a a^\dagger \rho_{ca} -2 a^\dagger \rho_{ca} a 
+ \rho_{ca} a a^\dagger\right)
\notag \\
&&
- \frac{r_2}{2} \phi^2 \left(a^\dagger a \rho_{ca} -2 a \rho_{ca} a^\dagger
+ \rho_{ca} a^\dagger a \right),
\label{Eq3}
\end{eqnarray}

\noindent
where $\phi = g \tau$ with $\tau$ the time spent by  atoms
inside the cavity. The coefficients $r_1$ and $r_2$ are the arrival rates for atoms being 
respectively in the excited and ground state. They  are related to the probabilities of 
emission  and absorption of a photon in the cavity and are explicitly given by
$r_1= \left(\rho_e + \rho_d + \rho_{nd} \right)$ and 
$r_2= \left(\rho_g + \rho_d + \rho_{nd} \right)$.
The asymptotic steady state solution of the master equation \eqref{Eq3} is the 
{\it thermal} distribution,
\begin{eqnarray}
\rho_{ca}^{s}
= \left(1 - e^{-\beta_{ca} \hbar \omega} \right) 
e^{-\beta_{ca} \hbar \omega a^\dagger a} \ .
\label{Eq4}
\end{eqnarray}
The inverse temperature of the cavity, $\beta_{ca} = (kT_{ca})^{-1}$, is determined by 
the coefficients $r_1$ and $r_2$ through the detailed--balance condition, 
$\exp(-\beta_{ca} \omega) = r_1/r_2$. When the interaction parameter $\lambda$ is zero,
the temperature of the cavity is equal to the atomic temperature, $T_{ca}=T$. On the other 
hand, when 
$\lambda\neq 0$, the diagonal and  nondiagonal matrix elements, $\rho_{d}$ and $\rho_{nd}$, 
modify  detailed--balance  between emission and absorption. As a result, the temperature of 
the cavity is effectively changed. 
 We will see below that $\lambda \neq 0$ indicates the 
presence of quantum correlations between two paired atoms.

The temperature of the cavity can be readily expressed in terms of the frequency 
$\omega$ of the two-level atoms and the interaction parameter $\lambda$. We have,

\begin{eqnarray}
\frac{\beta_{ca}}{\beta} 
= 1 - \frac{1}{\beta \hbar \omega}
\ln \left(
\frac{1 + e^{\beta \hbar \omega} e^{-\beta \hbar \lambda}}
{e^{\beta \hbar \omega} + e^{-\beta \hbar \lambda}} \right).
\label{Eq5}
\end{eqnarray}
The ratio $\beta_{ca}/\beta$ is always larger or equal
than one, implying that the temperature of the cavity is smaller than  the temperature 
of the reservoir when $\lambda \neq 0$.

\section{One atom through cavity}

We next consider the case where only one atom of the thermal entangled pair is sent 
through the cavity. The density operator of that atom is obtained from the total 
density matrix \eqref{Eq2} by tracing over the second atom. Due to the symmetry of 
the mixture, the two reduced operators are identical. 
By repeating the previous analysis, we find that the stationary state of the cavity 
field is thermally distributed with an inverse temperature, $\beta'_{ca}=(kT'_{ca})^{-1}$, 
satisfying  $\exp(-\beta'_{ca} \omega) = r'_1/r'_2$, with $r_1^{'} = \rho_e + \rho_d$ 
and $r_2^{'} = \rho_g + \rho_d$. Here, the coefficients $r'_1$ and $r'_2$ depend on the 
interaction parameter $\lambda$ only through the diagonal matrix elements $\rho_d$, since 
the nondiagonal elements do not appear. For the inverse temperature  of the cavity, we obtain,

\begin{eqnarray}
\frac{\beta'_{ca}}{\beta} 
=
1 - \frac{1}{\beta \hbar \omega} 
\ln{\frac{1 + e^{\beta \hbar \omega} \cosh(\beta \hbar \lambda)}
{e^{\beta \hbar \omega} + \cosh(\beta \hbar \lambda)}}.
\label{Eq6}
\end{eqnarray}
Contrary to the two--atom case, we observe that the inverse temperature ratio 
$\beta'_{ca}/\beta$ is  smaller or equal than one, indicating that the temperature 
of the cavity is larger than $T$ when $\lambda \neq 0$.

\section{Absorption and emission}

\begin{figure}
\center
\epsfxsize=0.48\textwidth
\epsffile{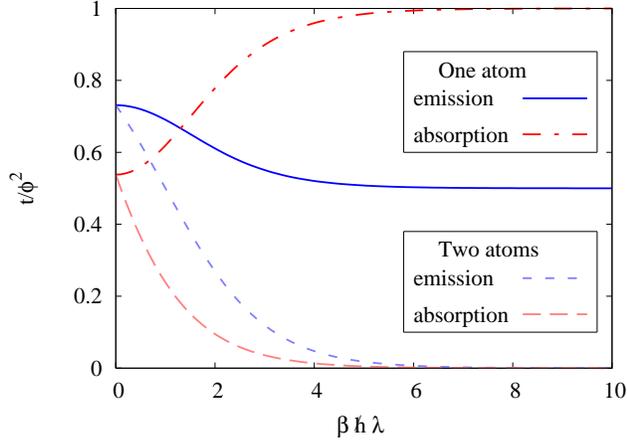}
\caption{Probabilities of absorption and emission, $t_{\pm}$, of a photon by the cavity mode, 
Eqs.~\eqref{eq8} and  \eqref{eq9}, from the oscillator state $n=1$, as a function of the 
interaction parameter $\beta \hbar \lambda$ for a fixed value of the temperature 
$\beta \hbar \omega=1$.}
\label{Fig1}
\end{figure}

Deeper insight into the previous results can be gained by directly examining 
the probabilities of  absorption and emission of a photon by the cavity mode. 
The latter can be easily calculated from the master equation  \eqref{Eq2} by 
evaluating the transition rates per unit time, $t_{\pm}$, from an oscillator 
state $|n \ra$ to the adjacent states $|n\pm 1\ra$ \cite{coh92}. They are given by

\bea
t_+ &=& r_1 \phi^2 \,(n+1) \ , \hspace{.6cm} \mbox{(absorption)}\label{eq8}\\
t_- &=& r_2 \phi^2 \,n \ ,\hspace{1.5cm} \mbox{(emission)\label{eq9}} 
\eea

\noindent
and similar expressions for the transition rates, $t'_{\pm}$, in  the one--atom case. 
The dependence of the different rates on the interaction parameter $\lambda$ is shown 
in Fig.~\ref{Fig1}. When a beam of correlated atomic  pairs is sent through the cavity, 
a larger value of $\lambda$ leads to a stronger reduction of  the  absorption rate than 
the emission rate. As a consequence, less energy is deposited in the cavity and its 
temperature is therefore effectively lowered. On the other hand, when only one atom of 
the pair passes through the cavity, the absorption probability is enhanced, while 
emission is suppressed. In contrast to the previous situation, more energy is thus 
provided to the cavity. This results in  an effective temperature rise.

The above results seem to contradict the zeroth law of thermodynamics which states that two objects in thermal equilibrium have the same temperature; one would then expect the temperature of the system to be equal to the temperature of the reservoir \cite{cal}.  The key observation to resolve this apparent  paradox is that, contrary to the usual situation in macroscopic thermodynamics,  here the system (cavity mode) does not interact with the reservoir (atomic beam) as a whole, but sequentially with its individual parts (atoms). Let us consider the total density operator of a pair of atoms as given by Eq.~\eqref{Eq2}. The latter corresponds to the equilibrium Gibbs state $\rho = Z^{-1} \exp(-\beta H)$. In the absence of correlations, the total density operator of the pair is the direct product, $\rho = \rho_{\mathcal{A}_1} \otimes \rho_{\mathcal{A}_2}$, of the reduced density operators of the individual atoms,   $\rho_{\mathcal{A}_i}
 =\tr_{\mathcal{A}_j} \rho$; they are also given by Gibbs states, $\rho_{\mathcal{A}_i} = Z_i^{-1} \exp(-\beta H_i)$, since $\lambda =0$. In the presence of correlations, however, the total state does not factorize and the density operators of     atoms 
$\mathcal{A}_1$ and $\mathcal{A}_2$  are no longer of the Gibbs form. Therefore, although the pair as a whole is in an equilibrium state, the individual atoms of the pair, with which the cavity mode interacts, are not.  The reduced density operators of the atoms are explicitly given by, 

\begin{eqnarray}
\rho_{\mathcal{A}_1} = \rho_{\mathcal{A}_2}  
= r_2 |g\rangle \langle g| + r_1 |e\rangle \langle e|.
\end{eqnarray}
 The level occupancies of  ground and excited states take their equilibrium values, $e^{\beta \hbar \omega}/2 \cos(\beta \hbar \omega)$ and 
$e^{-\beta \hbar \omega}/2 \cos(\beta \hbar \omega)$, only when $\lambda=0$. The resonant interaction of the cavity mode with the nonequilibium atoms then leads to the thermal state, Eq.~\eqref{Eq4}, with a temperature different from the reservoir temperature. This unusual feature is a consequence of the microscopic size of the photo--Carnot engine; it occurs for the quantum correlated as well as for the quantum coherent atomic reservoir \cite{Scully}. It is important to emphasize that the steady--state density operator of the cavity mode, Eq.~\eqref{Eq4}, is always given by a {\it thermal} state with a well-defined temperature for all values of the interaction parameter $\lambda$.

Interestingly, the temperature of the cavity differs from the atomic temperature 
in both cases, even when  $\beta \hbar \lambda \le \arcsinh(1)$, that is, even when 
the two--atom state is separable. This result shows that entanglement by itself is not 
responsible for the observed deviations. In the following, we express the modified 
cavity temperature in terms of the quantum discord. 

\section{Quantum discord}

The quantum discord is defined as the difference between two expressions of the 
mutual information that are classically equal \cite{OllivierZurek},

\begin{eqnarray}
\delta(1|2) = 
\mathcal{I}(1:2)
-\mathcal{J}(1:2) \ .
\label{10}
\label{Eq10}
\end{eqnarray}
The quantum mutual information $\mathcal{I}(1:2)$ is given by 
$\mathcal{I}(1:2) = H(1) + H(2) - H(1,2)$. Here $H(j) = - 
\tr \rho_{j} \log_2 \rho_{j}$, is the von Neumann entropy of 
the reduced density operator of atom $j$, and $H(1,2) = - \tr \rho \log_2 \rho$ 
is the joint entropy of  the total density matrix \eqref{Eq2}. The alternative 
quantum generalization of the mutual information is 
$\mathcal{J}(1:2) = H(1) - H(1| \left\{\Pi_{2,l} \right\})$; it describes 
the information gained about atom 1 as a result of a set of measurement 
$\left\{\Pi_{2,l} \right\})$ on atom 2.
The conditional entropy, $H(1|\left\{\Pi_{2,l} \right\})= \sum_l p_l  H(1|\Pi_{2,l})$, is 
the sum  of the  von Neumann entropies $H(1|\Pi_{2,l})$ of the conditional density operators, 
$\rho_{1|\Pi_{2,l}} =  \Pi_{2,l} \rho \Pi_{2,l}/p_l$, after a perfect measurement of atom $2$. 
The operators  $\{\Pi_{2,l}\}$ define a set of orthogonal projectors on atom $2$ 
and $p_l = \tr_{1,2} \Pi_{2,l} \rho$ denotes the probability of occurrence of 
state $\rho_{1|\Pi_{2,l}}$ in a given measurement. The mutual information 
$\mathcal{J}(1:2)$ usually depends on the measurement basis and the discord has 
to be minimized over all sets $\{\Pi_{2l}\}$. Using the parametrized basis, 
$\{ \cos \theta |g\ra -\sin \theta |e\ra, -\sin \theta |g\ra -\cos \theta |e\ra\}$, 
we find the minimum of the discord for $\theta = \pi/4$:

\begin{eqnarray}
\delta(1|2) = -\frac{1}{\ln(2)} \Big[
2 \left( \beta \hbar \lambda \right)\rho_{nd} 
+\sum_{\alpha = g,e} \rho_\alpha 
\ln \left(\frac{\rho_\alpha + \rho_d}{\rho_\alpha} \right)
\notag \\
+\rho_d \ln \left(Z^2 \left(\rho_g+\rho_d \right)\left(\rho_e+\rho_d \right)\right)
+\sum_{\varepsilon = \pm} \Phi_\varepsilon \ln \Phi_\varepsilon
\Big],
\notag \\
\label{Eq18}
\end{eqnarray}
where we have defined $\Phi_\varepsilon =
(1 + \varepsilon \sqrt{\left(\rho_e-\rho_g \right)^2+4 \rho_{nd}^2}) /2$.
The discord vanishes when the interaction parameter $\lambda$ is zero, corresponding 
to the absence of quantum correlations between the two atoms. For maximally correlated 
states, the discord is unity and is, in this particular case, equal to the concurrence 
$\mathcal{C}$. In general, however, the discord is different from the concurrence and 
can even be nonzero when the state is separable.

\begin{figure}
\center
\epsfxsize=0.47\textwidth
\epsffile{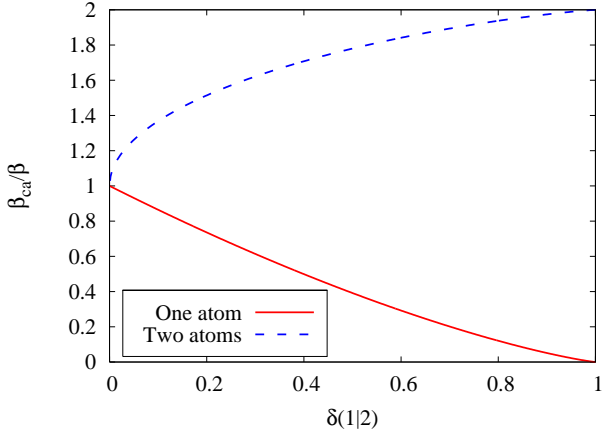}
\caption{ Relative inverse temperature of the cavity, $\beta_{ca}/\beta$, 
Eqs.~\eqref{Eq5} and \eqref{Eq6}, as a function of the quantum discord 
$\delta(1|2)$, Eq.~\eqref{Eq18},  for the parameter $\beta \hbar \omega = 10^{-3}$. 
The full (dashed) line corresponds to the situation in which one (two) 
atom(s) of the thermal entangled pair goes through the 
optical cavity.}
\label{Fig2}
\end{figure}

Figure \ref{Fig2} shows the ratio of the temperature of the cavity and the 
reservoir temperature, Eqs.~\eqref{Eq5} and \eqref{Eq6}, as a function of the 
quantum discord \eqref{Eq18} for a fixed value of $\beta \hbar \omega$. We first 
observe that the temperature of the cavity is equal to the reservoir temperature in 
the absence of quantum correlations when the discord is zero. We further clearly 
recognize that the temperature of the cavity increases with the discord in the case 
where only one of the two correlated atoms passes through the cavity, and decreases 
when both atoms fly through the cavity. We can therefore conclude that the deviation 
of the cavity temperature from the reservoir temperature  is induced by the quantum 
correlations of the thermal entangled pair. The maximum temperature deviation is 
obtained for maximal quantum correlations, $\delta(1|2) =1$.

For general correlated states,  quantum and classical correlations  coexist and   quantifying them is a nontrivial task. 
Following Henderson and Vedral \cite{hen01}, we compute
classical correlations in a  quantum system as the maximum 
over all measurements of the quantum mutual information $\mathcal{J}(1:2)$,

\begin{eqnarray}
C(1|2) = \mbox{max}_{\{\Pi'_{2,l}\}} \mathcal{J}(1:2).
\label{eq12}
\end{eqnarray}
For the  bipartite thermal entangled state, Eq.~\eqref{Eq4},  we find that the set of measurements that 
minimizes the quantum discord also maximizes classical correlations. In view of the 
definition of these two quantities, we thus obtain that $\mathcal{I}(1:2) = 
\delta(1|2)+ C(1|2)$. In other words, classical and quantum correlations do add up in this case to the total correlations, as given by the mutual information.  We have verified that this result also holds for generalized 
POVM measurements \cite{ham04}. 

\begin{figure}
\center
\epsfig{file=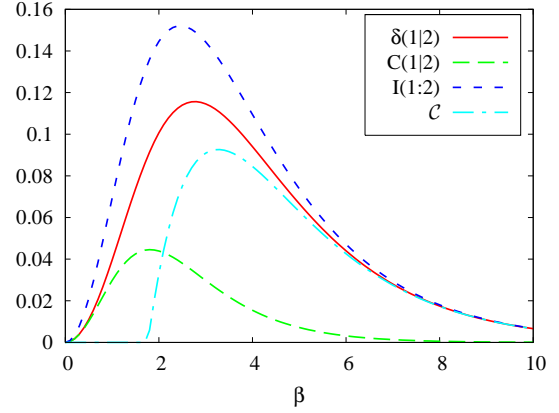,width=8cm}
\caption{
Four measures of the correlations in the thermal entangled state $\rho$ as a function 
of the inverse temperature $\beta$: quantum discord $\delta(1|2)$, classical correlations 
$C(1|2)$, total correlations $\mathcal{I}(1:2)$ and concurrence ${\cal C}$, for 
$\hbar \omega=1$ and $\hbar \lambda= 0.5$. For high temperatures, the concurrence 
is zero and classical correlations are equal to quantum correlations. By contrast, 
for low temperatures, classical correlations vanish and $\mathcal{I}(1:2) \simeq 
\delta(1|2) \simeq \mathcal{C}$.}
\label{FigSpecialConditions}
\end{figure}

The inverse temperature dependence of the four correlation measures, $\mathcal{I}(1:2)$,  
$\delta(1|2)$, $C(1|2)$ and ${\cal C}$, is plotted in Fig.\ref{FigSpecialConditions} 
for a fixed value of the interaction parameter $\lambda<\omega$. In the limit of 
high temperatures, $ \beta \hbar \omega \ll 1$, the atomic pair is separable (${\cal C}=0)$  
and quantum discord is equal to classical correlations, $\delta(1|2) \simeq C(1|2)$. 
It should be stressed that classical correlations are in general much smaller than quantum 
correlations. For small interaction parameter, $\beta \hbar \lambda\ll1$, Eqs.~\eqref{Eq18} 
and \eqref{eq12} simplify to $ \delta(1|2) \simeq C(1|2) \simeq (\beta \hbar \lambda)^2/8\ln 2$. 
Hence, for the atomic pairs passing through the cavity, we can directly express the effective 
temperature of the cavity in terms of 
the quantum mutual information using Eq.~\eqref{Eq5}, yielding a square--root 
dependence, $T_{ca}=T(1-\sqrt{{\cal I}(1:2) \ln2})$. On the other hand, for low temperatures, 
classical correlations vanish, as expected, and mutual information, discord and concurrence 
become identical. In this regime, we find that $\mathcal{I}(1:2) \simeq \delta(1|2) 
\simeq {\cal C} \simeq \exp(\beta \hbar(\lambda-\omega)) $. It then follows that, for 
small interaction parameter,  $T_{ca} = -T (\beta\hbar \omega)^{-1} \ln\mathcal{I}(1:2)$.

\section{Thermodynamic efficiency}

We are now in the position to show that quantum correlations can increase the thermodynamic efficiency above its classical value.
We consider a usual Carnot cycle where the cavity is successively put into 
contact with a high and low temperature reservoir as discussed in detail in Ref.~\cite{Scully}. 
In the situation where the two atoms of the pairs pass through the cavity, 
the atomic reservoir is taken to be the cold reservoir, while the opposite choice is made for the one--atom case. For high temperatures, taking into account that $\mathcal{I}(1:2) 
= \delta(1|2)+ C(1|2)$, the thermodynamic efficiency can be written as,

\begin{eqnarray}
\eta =\eta_C + \frac{T_c}{T_h} \, \sqrt{(\delta(1|2)+C(1|2))\ln2} \ .
\label{Eq22}
\end{eqnarray}

\noindent 
The efficiency $\eta$ can thus be decomposed into the classical efficiency 
$\eta_C$ (without correlations), and the contributions from the quantum correlations $\delta(1|2)$ and 
the accompanying classical correlations $C(1|2)$. 
By contrast, for lower temperatures, since classical correlations vanish and $\mathcal{I}(1:2) 
\simeq \delta(1|2) \simeq {\cal C}$, we can directly relate the efficiency 
to the concurrence of an  entangled pair,

\begin{eqnarray}
\eta =\eta_C + \frac{T_c}{T_h}\left(1 + \frac{1}{\beta \hbar \omega} \ln\mathcal{C}\right) \ .
\label{Eq23}
\end{eqnarray}

\noindent
The above equations show that the thermodynamic  efficiency in presence of correlations between the 
two atoms exceeds the classical efficiency $\eta_{C}$ correspondig to uncorrelated atoms.
Furthermore, we note that the quantum heat engine can still extract work from the 
external reservoirs when their temperatures are equal, $T_c=T_h$ \cite{Scully}.
It should be emphasized that the second law of thermodynamics is not violated even 
though work can be gained from a single reservoir. Indeed the interaction between  atoms and   cavity field  induces a temperature difference between the two only when  atomic pairs are correlated, and   work has to be provided to prepare the correlated thermal entangled state \eqref{Eq2}. 
Finally, it is interesting to mention that a high--quality optical cavity with a movable mirror has recently been fabricated with a micromechanical mirror \cite{boh06}.

\section{Conclusion}

We have shown that a beam of quantum correlated two--level atoms can modify 
detailed--balance between absorption and emission of photons in a single--mode cavity and 
change its effective temperature. In the limit of small correlations, we have expressed the  efficiency the microscopic photo--Carnot engine  in terms of the quantum discord and   shown that the efficiency exceeds the classical limit given by uncorrelated atoms.
Useful work can thus be extracted from quantum correlated, but not necessarily entangled, thermal qubits.
These findings show that quantum correlations cannot only be used as a valuable 
resource in quantum information theory, but also in the realm of quantum thermodynamics. 

\acknowledgements{This work was supported by  the
Emmy Noether Program of the DFG (contract LU1382/1-1) and the
cluster of excellence Nanosystems Initiative Munich (NIM).}

\end{document}